\documentclass[prd,12pt,tightenlines,showpacs,amsmath,amssymb]{revtex4}
\usepackage{bm}

\DeclareFontFamily{OT1}{rsfs}{}
\DeclareFontShape{OT1}{rsfs}{m}{n}{
                 <-7> rsfs5 <7-10> rsfs7 <10-> rsfs10}{}
\DeclareMathAlphabet{\mycal}{OT1}{rsfs}{m}{n}

\begin{document}

\preprint{gr-qc/0212050}

\title{``No--Scalar--Hair'' Theorems for Nonminimally Coupled
Fields with Quartic Self--Interaction}

\author{Eloy Ay\'on--Beato}
 \email{ayon@fis.cinvestav.mx}
\affiliation{Departamento~de~F\'{\i}sica,%
~Centro~de~Investigaci\'on~y~de~Estudios~Avanzados~del~IPN,\\
Apdo.~Postal~14--740,~07000,~M\'exico~D.F.,~M\'exico.}

\begin{abstract}
Self--gravitating scalar fields with nonminimal coupling to
gravity and having a quartic self--interaction are considered in
the domain of outer communications of a static black hole. It is
shown that there is no value of the nonminimal coupling parameter
$\zeta$ for which nontrivial static black hole solutions exist.
This result establishes the correctness of Bekenstein
``no--scalar--hair'' conjecture for quartic self--interactions.
\end{abstract}


\maketitle

\section{Introduction}

Black holes have no ``hair'' is the classical statement which
summarizes the physics behind stationary configurations describing
the final state of highly massive matter under gravitational
collapse. The ``no--hair'' conjecture stresses the fact that
short--range interactions decaying fast enough at infinity
(``hair''), so that they have no contribution to Gauss--like
conserved charges, are not allowed in the exterior of a stationary
black hole. Hence, the black--hole classical degrees of freedom
are restricted to those related to its conserved charges.

The accuracy of this conjecture was challenged in the last decade
when several ``hairy'' black hole solutions were discovered (see
Ref.~\cite{Bizon94} for a review and for the latest results see
Ref.~\cite{Galtsov01} and references therein). All these black
holes shared together the feature of having nonlinear matter
fields with no associated conserved charges.

It is believed that the conjecture remains true without reserve
for other nonlinear fields. For example, a particular case of the
conjecture concerning merely scalar fields has been formulated by
Bekenstein and investigated by him and many other authors. The
``no--scalar--hair'' conjecture~\cite{Bekens96} asserts that for
neutral scalar fields, nonminimally coupled to gravity, and with
self--interaction potentials nontrivial black hole solutions are
forbidden, i.e., any stationary black hole solution of the system
must has a trivial scalar field (zero or constant) and
consequently the solution must be given by one of the well--known
vacuum black holes (Schwarzschild or Kerr). The aim of this work
is to settle this conjecture for quartic self--interactions, i.e.,
to exclude static black hole hairs in the framework of
nonminimally coupled to gravity quartic self--interacting scalar
fields.

Scalar fields with minimal coupling and a quartic
self--interaction were earlier studied on the background of
Schwarzschild solution, exhibiting that only nonregular field
configurations are allowed~\cite{Sawyer77,Brum78}. Indeed, the
present understanding of the topic excludes the existence of
self--gravitating spherical scalar hairs with a general
nonnegative self--interaction potential under minimal
coupling~\cite{HeuslerStrau92,Heusler92,Sudarsky95,Heusler95,Bekens95}.

Comprising nonminimal couplings and nonnegative
self--interactions, hairy spherical black holes are inadmissible
for nonminimal coupling parameters in the ranges $\zeta<0$ and
$\zeta\geq1/2$~\cite{MayoBeke96,Bekens96}. Other restrictions on
the nonminimal couplings have been established also in $n>3$
dimensions~\cite{SaaPRD96}, and even for
$(2+1)$--gravity~\cite{AyonGMP00}. More recently, numerical
evidence against the existence of nontrivial black hole solutions
for each value of the nonminimal coupling spectrum under
consideration has been presented in Ref.~\cite{PenaSudarsky01}.

In this work we analyze nonminimally coupled to gravity scalar
fields with quartic self--interactions. The motivation for
choosing a quartic self--interaction rests on the following: for
massless scalar fields, several works on the
subject~\cite{XanZan91,XanDia92,Klimcik93,Zannias95} reinforce the
idea on the uniqueness of the nontrivial conformal Bekenstein
black hole~\cite{BBM70,Bekens74}. Independently of a recent
controversial debate on the existence of this black hole as a
conventional exact solution to Einstein
equations~\cite{SudaZann98,Bekens98}, our view is that the
oddities concerned with this solution may be due to the conformal
invariance of the related model---actually, this solution was
originally generated by a conformal
transformation~\cite{Bekens74}. On the other hand, it is known
that in four dimensions the relevant model remains conformally
invariant under the inclusion of a quartic self--interaction.
Hence, we decide to explore if conformal invariance also implies
in this new case a nontrivial behavior in the presence of static
black holes. Self--gravitation of these systems, including not
only conformal coupling, is described by the action
\begin{equation}
S=\frac12\int dv\left( \frac{1}{\kappa}R
-\nabla_\mu\Phi\nabla^\mu\Phi - \frac12\lambda\Phi^4
-\zeta\,R\,\Phi^2 \right), \label{eq:ac}
\end{equation}
where $R$ stands for the scalar curvature, $\Phi$ is the scalar
field, $\zeta$ is a parameter describing the nonminimal coupling
between gravity and the scalar field, and $\lambda$ is a
self--coupling parameter characterizing the quartic
self--interaction of the scalar field. We shall consider all the
real values of the nonminimal coupling parameter $\zeta$, on the
other side the parameter $\lambda$ is taken positive in order to
the potential be nonnegative; a required condition for ruled out
scalar hairs in the minimal model ($\zeta=0$)~\cite{Heusler95}.

The ``no--scalar--hair'' theorem we attempt to establish in this
work consists in demonstrating that for all values of the
nonminimal coupling parameter $\zeta$, including the conformal one
($\zeta=1/6$), any nontrivial behavior of the scalar field in
presence of static asymptotically flat black holes is completely
forbidden. That is, we shall study the scalar equation
\begin{equation}
\Box\Phi - \lambda\Phi^3 -\zeta\,R\,\Phi = 0,
\label{eq:K-G}
\end{equation}
together with the Einstein equations
\begin{equation}
\left(1-\kappa\zeta\Phi^2\right)R_\nu^{~\mu}
=\kappa\left[\nabla_\nu\Phi\nabla^\mu\Phi
+\frac{1}{2}\delta_\nu^{~\mu}
 \left(\frac12\lambda\Phi^4-\zeta\Box\Phi^2\right)
-\zeta\nabla_\nu\nabla^\mu\Phi^2
       \right],
\label{eq:Ein}
\end{equation}
arising both from action (\ref{eq:ac}), not only in the case of
the conformal coupling but for all the couplings permitted by the
values of $\zeta$, and we shall show that a scalar field
fulfilling the above equations must vanish in the domain of outer
communications of a static black hole.

In a static black hole, the stationary Killing field $\bm{k}$
coincides with the null generator of the event horizon
$\mycal{H}^+$ in addition to being timelike and hypersurface
orthogonal in all the domain of outer communications
$\ll\!\!\mycal{I}\!\!\gg$ (see Ref.~\cite{Ayon01} for sufficient
conditions allowing that a stationary spacetime containing the
above matter can be static). The simply connectedness of
$\ll\!\!\mycal{I}\!\!\gg$ \cite{ChrWald95,Galloway95}, together
with the above features, implies the existence of a global
coordinate system, $(t,x^i)$, $i=1,2,3$, in
$\ll\!\!\mycal{I}\!\!\gg$ \cite{Carter87} where
$\bm{k}=\bm{\partial/\partial t}$ and the metric can be expressed
by
\begin{equation}
\bm{g}=-V\bm{dt}^2+\gamma_{ij}\bm{dx}^i\bm{dx}^j,
\label{eq:static}
\end{equation}
where $V$ and $\bm{\gamma}$ are $t$--independent, $\bm{\gamma}$ is
positive definite in all of $\ll\!\!\mycal{I}\!\!\gg$, and $V$ is
a positive function tending to zero on $\mycal{H}^+$.

In the following section, it is shown that, in the domain of outer
communications of a static asymptotically flat black hole,
self--gravitating scalar fields with quartic self--interaction and
nonminimally coupled to gravity, with nonminimal coupling
parameter satisfying $\zeta<0$ or $\zeta>1/6$ vanish. Section
\ref{sec:equ} is devoted to establishing the result for $\zeta=0$
and $\zeta=1/6$, i.e., the minimal and conformal couplings
respectively. In Sec.~\ref{sec:great} a similar result is
established for the remaining values of the nonminimal coupling
parameter within the interval $0<\zeta<1/6$. Finally, we provide
some conclusions in the last Sec.~\ref{sec:conclu}.

\section{``No--Scalar--Hair'' Theorem for $\zeta<0$ and
$\zeta>1/6$ Couplings}
\label{sec:les}

In this section we analyze simultaneously the nonminimal couplings
with $\zeta<0$ as well as $\zeta>1/6$; in both cases, it is
possible to make the following redefinition for the scalar field
$\varphi\equiv
\left[6\,\kappa\,\zeta\left(\zeta-1/6\right)\right]^{1/2}\Phi$.
Upon contracting the Einstein equations (\ref{eq:Ein}) and
inserting the scalar field equation (\ref{eq:K-G}) in the result,
the following expression for the scalar curvature, in terms of the
redefined scalar field, arises
\begin{equation}
-\zeta{R}=\frac{\nabla_\mu\varphi\nabla^\mu\varphi}{1+\varphi^2}
+\lambda{b}^2\frac{\varphi^4}{1+\varphi^2} , \label{eq:curvII}
\end{equation}
where
$b\equiv\left[6\,\kappa\,\zeta\left(\zeta-1/6\right)\right]^{-1/2}$.
The right--hand side of this expression contains nonnegative
terms. In particular, the kinetic term can be written as
$\nabla_\mu\varphi\nabla^\mu\varphi
=\gamma_{ij}\nabla^i\varphi\nabla^j\varphi$ in the coordinates
used in (\ref{eq:static}), where $\bm{\gamma}$ is positive
definite. Two facts follow from this consideration, first, by
asymptotic flatness ($R=0$) the scalar field vanishes
asymptotically due to vanishing of the second term at the
right--hand side of (\ref{eq:curvII}). Second, since the domain of
outer communications and the event horizon are regular regions,
the left--hand side of (\ref{eq:curvII}) is bounded everywhere,
and hence, from the bounded behavior of the second term on the
right--hand side of (\ref{eq:curvII}), the scalar field is also
bounded everywhere.

Inserting now the expression for $R$ (\ref{eq:curvII}) in the
scalar equation (\ref{eq:K-G}), in terms of $\varphi$, one arrives
at
\begin{equation}
\Box\varphi=\frac{1}{1+\varphi^2}\left(
-\varphi\nabla_\mu\varphi\nabla^\mu\varphi
+\lambda{b}^2\varphi^3\right) .
\label{eq:scalarII}
\end{equation}
By means of the above equation, we shall show  that $\varphi$
vanishes in all of $\ll\!\!\mycal{I}\!\!\gg$, for a static
asymptotically flat black hole.

Let $\mycal{V}\subset\ll\!\!\mycal{I}\!\!\gg$ be the open region
bounded by the spacelike hypersurface $\Sigma$, the spacelike
hypersurface $\Sigma^{\prime}$, and the corresponding portions of
the horizon $\mycal{H}^+$ and the spatial infinity $i^o$; where
the spacelike hypersurface $\Sigma^{\prime}$ is obtained by
shifting each point of $\Sigma$ by a unit value of the parameter
along the integral curves of the stationary Killing field
$\bm{k}$.

Multiplying the scalar equation (\ref{eq:scalarII}) by $\varphi$,
integrating by parts over $\mycal{V}$, and using the Gauss
theorem, one arrives at the following integral identity,
\begin{equation}
\left[\int_{\Sigma^{\prime}} -\int_\Sigma +\int_{\mycal{H}^+\cap
\overline{\mycal{V}}} +\int_{i^o\cap\overline{\mycal{V}}}\right]
\varphi\nabla^\mu\varphi\,d\Sigma_\mu=\int_\mycal{V}\left(
\frac{\nabla_\mu\varphi\nabla^\mu\varphi}{1+\varphi^2}
+\lambda{b}^2\frac{\varphi^4}{1+\varphi^2}\right) dv,
\label{eq:intII}
\end{equation}
where $\overline{\mycal{V}}$ stands for the closure of the set
$\mycal{V}$. On the left--hand side of (\ref{eq:intII}) the
boundary integral over the hypersurface $\Sigma^{\prime}$ cancels
out that over $\Sigma$, since $\Sigma^{\prime}$ and $\Sigma$ are
isometric surfaces having inverted normals.

By asymptotic flatness ($R=0$) the boundary integral over
$i^o\cap\overline{\mycal{V}}$ vanishes, since $\varphi$ approaches
asymptotically the zero value, see expression (\ref{eq:curvII}).
In fact, for a stationary asymptotically flat spacetime scalar
curvature behaves at infinity as $R=O(1/r^3)$, where
$r^2\equiv\sum_{i=1}^{3}(x^i)^2$ is the radial coordinate related
to asymptotically Euclidean coordinates at infinity. From relation
(\ref{eq:curvII}) between the scalar curvature and the scalar
field it follows that $\varphi=O(1/r^{3/4})$ at infinity in order
that the existence of the scalar field is consequent with
asymptotic flatness. Hence, the boundary integrand behaves as
$\varphi\nabla_\mu\varphi=O(1/r^{5/2})$ and since the surface
element goes as $d\Sigma^\mu=O(r^2)$, it can be concluded that the
boundary integral vanishes appropriately at infinity.

We now shall prove that the boundary integral over the portion of
the horizon $\mycal{H}^+\cap\overline{\mycal{V}}$ also vanishes.
To this end, we select the most natural volume 3--form at the
horizon which is given by $\bm{\eta_3}=-*\!\bm{n}$, where, if
$\bm{l}$ is the null generator of the horizon, $\bm{n}$ is the
other future--directed null vector ($n_{\mu}l^\mu=-1$) orthogonal
to the spacelike cross sections of the horizon, and the star $*$
stands for the Hodge dual operator. It can be shown that with this
volume 3--form the measure of integration on $\mycal{H}^+$ is the
standard one in this region~\cite{Zannias95,Zannias98,Ayon02a}
\[
d\Sigma_\mu=2n_{[\mu}l_{\nu]}l^{\nu}d\sigma ,
\]
where $d\sigma$ is the surface element. With the above measure,
the integrand at the horizon is written as
\begin{equation}
\varphi\nabla^\mu\varphi\,d\Sigma_\mu
=\varphi\left[l_\mu\nabla^\mu\varphi
+\left(l_\nu{l}^\nu\right)
n_\mu\nabla^\mu\varphi\right]d\sigma.
\label{eq:hintef}
\end{equation}
The stationary Killing field $\bm{k}$ of a static black hole
coincides at the horizon with its null generator $\bm{l}$, hence,
$l_\mu\nabla^\mu\varphi=\bm{\pounds_k}\varphi=0$ by the staticity
of the scalar field $\varphi$. Then, using also the bounded
behavior of $\varphi$ (\ref{eq:curvII}), the first term of the
integrand (\ref{eq:hintef}), $\varphi\,l_\mu\nabla^\mu\varphi$,
must vanish at the horizon. To show the vanishing at the horizon
of the second term of the integrand (\ref{eq:hintef}),
$\varphi\left(l_\nu{l}^\nu\right) n_\mu\nabla^\mu\varphi$, we
shall introduce a null tetrad basis at the horizon (see also the
approach of~\cite{Zannias95,Zannias98} on this subject). This
tetrad is composed by the null fields $\bm{l}$ and $\bm{n}$,
together with two linearly independent spacelike fields tangent to
the spacelike cross sections of the horizon. Using this tetrad the
metric is written at the horizon as
\begin{equation}
\bm{g}=-\bm{l}\otimes\bm{n} - \bm{n}\otimes\bm{l} + \bm{g}^\bot,
\label{eq:hmetric}
\end{equation}
where $\bm{g}^\bot$ stands for the metric projection on the
spacelike cross sections of the horizon, which are orthogonal by
definition to the null vectors $\bm{l}$ and $\bm{n}$. Expanding
the gradient of the scalar field $\bm{\nabla}\varphi$ in the above
base at the horizon, we obtain
\begin{equation}
\bm{\nabla}\varphi= -\left(n_\mu\nabla^\mu\varphi\right)
\bigg{[}\bm{l} +\left(l_\nu{l}^\nu\right)\bm{n}\bigg{]}
+\bm{\nabla}^\bot\varphi, \label{eq:Xexpan}
\end{equation}
where orthogonality of the scalar gradient to the null generator
of the horizon has been used, $l_\mu\nabla^\mu\varphi=0$. Here
again the notation $\bm{\nabla}^\bot$ represents the projection of
the gradient on the spacelike cross sections of the horizon, being
orthogonal to $\bm{l}$ and $\bm{n}$. Special attention must be
given to the term involving the norm of the null generator; one
has to keep it in the expansion, since there is no evidence
\emph{a priori} that this term vanishes. Using the expansion
(\ref{eq:Xexpan}), we write the square of the scalar field
gradient as
\begin{equation}
\nabla_\mu\varphi\nabla^\mu\varphi=
-\left(l_\nu{l}^\nu\right)\left(n_\mu\nabla^\mu\varphi\right)^2
+\nabla_\mu^\bot\varphi\nabla^{\bot\mu}\varphi.
\label{eq:squareD1}
\end{equation}
Inserting directly the expression of the metric in the given
tetrad (\ref{eq:hmetric}) within the square of the scalar field
gradient, a different representation for this kinetic term can be
obtained, i.e.,
\begin{eqnarray}
\nabla_\mu\varphi\nabla^\mu\varphi
&=&g_{\mu\nu}\nabla^\mu\varphi\nabla^\nu\varphi
\nonumber\\
&=&\left(-l_\mu{n}_\nu-n_\mu{l}_\nu+g^\bot_{\mu\nu}\right)
\nabla^\mu\varphi\nabla^\nu\varphi\nonumber\\
&=&-2\left(l_\nu\nabla^\nu\varphi\right)
\left(n_\mu\nabla^\mu\varphi\right)
+\nabla_\mu^\bot\varphi\nabla^{\bot\mu}\varphi.
\label{eq:squareD2}
\end{eqnarray}
Subtracting the expressions for the square of the gradient
(\ref{eq:squareD1}) and (\ref{eq:squareD2}), one obtains
\begin{equation}
\bigg{[}\left(l_\nu{l}^\nu\right)
\left(n_\mu\nabla^\mu\varphi\right)-
2\left(l_\mu\nabla^\mu\varphi\right)\bigg{]}
\left(n_\mu\nabla^\mu\varphi\right)=0.
\label{eq:2tzero}
\end{equation}
The case in which $n_\mu\nabla^\mu\varphi=0$ is trivial, since the
product of this term times the norm of the null generator,
$\left(l_\nu{l}^\nu\right)n_\mu\nabla^\mu\varphi$, automatically
vanishes as well. The interesting case occurs when
$n_\mu\nabla^\mu\varphi\neq0$, under this circumstance the term
within the brackets in (\ref{eq:2tzero}) must vanish, which in
turn implies, from the staticity of the scalar field
$l_\mu\nabla^\mu\varphi=\bm{\pounds_k}\varphi=0$, that the term
$\left(l_\nu{l}^\nu\right)n_\mu\nabla^\mu\varphi$ is vanishing.
From these properties, using the bounded behavior of the scalar
field (\ref{eq:curvII}), it follows that the second term of the
integrand (\ref{eq:hintef}),
$\varphi\left(l_\nu{l}^\nu\right)n_\mu\nabla^\mu\varphi$, also
vanishes at the horizon.

With vanishing of the boundary integral at the portion of the
horizon $\mycal{H}^+\cap\overline{\mycal{V}}$, there remain no
contributions on the left--hand side of (\ref{eq:intII}), thus
\begin{equation}
\int_\mycal{V}\left(
\frac{\gamma_{ij}\nabla^i\varphi\nabla^j\varphi}{1+\varphi^2}
+\lambda{b}^2\frac{\varphi^4}{1+\varphi^2}\right)dv=0,
\label{eq:zerosII}
\end{equation}
using the coordinates of (\ref{eq:static}). Since $\bm{\gamma}$ is
positive definite, the integrand above is non--negative, hence
(\ref{eq:zerosII}) implies vanishing of each term of this
integrand in $\mycal{V}$, from which it follows finally that
$\varphi$ vanishes in $\mycal{V}$, and therefore in all of
$\ll\!\!\mycal{I}\!\!\gg$ by staticity.

In this way, the ``no--scalar--hair'' theorem has been proven for
a field with a quartic self--interaction with $\zeta<0$ or
$\zeta>1/6$ nonminimal couplings to the gravity of a static
asymptotically flat black hole.

\section{``No--Scalar--Hair'' Theorem for $\zeta=0$ and
$\zeta=1/6$ Couplings}
\label{sec:equ}

This section deals with the minimal, $\zeta=0$, and conformal,
$\zeta=1/6$, couplings. For the minimal case the results of
references~\cite{HeuslerStrau92,Heusler92,Sudarsky95,Heusler95,Bekens95}
are known, which apply to any non--negative self--interaction. For
the conformal case there are still no results in the literature on
the subject using self--interactions, except those of
Ref.~\cite{AyonPerez01} for massive fields.

The reason for dealing with both couplings in the same fashion is
that the corresponding scalar equation is the same in both cases.
Actually, for the minimal coupling $\zeta=0$ the scalar equation
(\ref{eq:K-G}) reduces to
\begin{equation}
\Box\Phi -\lambda\Phi^3=0.
\label{eq:KGR0}
\end{equation}
The same equation also describes the conformal coupling
$\zeta=1/6$, due to the fact that the conformal symmetry implies
$R=0$; this can be noted from the contraction of the Einstein
equations (\ref{eq:Ein}) for $\zeta=1/6$, and the insertion of the
scalar equation (\ref{eq:K-G}) into the result.

The only constant solution to equation (\ref{eq:KGR0}) is the
trivial one: $\Phi=0$; hence, it is the only correct asymptotic
solution in an asymptotically flat spacetime. We shall show for
both couplings $\zeta=0$ and $\zeta=1/6$ that if the scalar field
$\Phi$ vanishes asymptotically, it must vanish in all of
$\ll\!\!\mycal{I}\!\!\gg$.

Multiplying equation (\ref{eq:KGR0}) by $\tanh\Phi$ and
integrating by parts over the previously defined region
$\mycal{V}$ gives, after using the Gauss theorem,
\begin{eqnarray}
\left[\int_{\Sigma^{\prime}} -\int_\Sigma +\int_{\mycal{H}^+\cap
\overline{\mycal{V}}} +\int_{i^o\cap\overline{\mycal{V}}}\right]
\tanh{\Phi}\,\nabla^\mu\Phi\,d\Sigma_\mu &&\nonumber\\
=\int_\mycal{V}\left( {\rm sech}^2{\Phi}\,
\nabla_\mu\Phi\nabla^\mu\Phi +\lambda\Phi^3\tanh{\Phi} \right)
dv.&& \label{eq:intIII}
\end{eqnarray}
On the left--hand side of (\ref{eq:intIII}), the integrals over
the hypersurfaces $\Sigma^{\prime}$ and $\Sigma$ mutually cancel
out again by isometry between both hypersurfaces, and by the
opposite direction of its normals in the boundary integration. The
integral over $i^o\cap\overline{\mycal{V}}$ is zero because the
scalar field $\Phi$ must vanish asymptotically (see the discussion
in~\cite{Zannias98} on the fall--off rates for the scalar field
following from the left--hand side of the Einstein equations in
Bondi coordinates, such arguments are independent of if there
exists a self--interaction potential or not). The remaining
boundary integral over the portion of the horizon
$\mycal{H}^+\cap\overline{\mycal{V}}$ has again a vanishing
integrand; using the same arguments as in the previous section it
can be shown that the term $\nabla^\mu\Phi\,d\Sigma_\mu$ of the
integrand vanishes at the horizon and since $\tanh\Phi$ is ever
bounded, independent of the value of the scalar field at the
horizon, the product of both terms,
$\tanh{\Phi}\,\nabla^\mu\Phi\,d\Sigma_\mu$, vanishes at the
horizon.

Therefore, the boundary integral in (\ref{eq:intIII}) vanishes
and, in the coordinates of (\ref{eq:static}), we obtain
\begin{equation}
\int_\mycal{V}\left( {\rm sech}^2{\Phi}\,
\gamma_{ij}\nabla^i\Phi\nabla^j\Phi +\lambda\Phi^3\tanh{\Phi}
\right) dv=0. \label{eq:zerosIII}
\end{equation}

The above integral identity, together with the non--negativeness
of the associated integrand terms within it, imply vanishing of
the scalar field $\Phi$ in $\mycal{V}$, and hence in all the
domain of outer communications $\ll\!\!\mycal{I}\!\!\gg$.

\section{``No--Scalar--Hair'' Theorem for $0<\zeta<1/6$ Couplings}
\label{sec:great}

At last, we shall concentrate on this section in the remaining
interval, $0<\zeta<1/6$, of nonminimal couplings to gravity. The
procedure is similar to that used in Sec.~\ref{sec:les}, but
demands a more careful treatment of these couplings. In this case
it will be also useful to redefine the scalar field,
$\varphi\equiv
\left[6\,\kappa\,\zeta\left(1/6-\zeta\right)\right]^{1/2}\Phi$
(notice the difference with the notation used in
Sec.~\ref{sec:les}).

The scalar curvature in the above interval of nonminimal
couplings, obtained by substitution of the scalar equation
(\ref{eq:K-G}) into the contracted Einstein equations
(\ref{eq:Ein}), and with the above redefinition of the scalar
field, is expressed by
\begin{equation}
\zeta{R}=\frac{\nabla_\mu\varphi\nabla^\mu\varphi}{1-\varphi^2}
+\lambda{b}^2\frac{\varphi^4}{1-\varphi^2} ,
\label{eq:curvIV}
\end{equation}
now
$b\equiv\left[6\,\kappa\,\zeta\left(1/6-\zeta\right)\right]^{-1/2}$.
Each term on the right--hand side of this expression has the same
sign again. Since the left--hand side of the above relation is
bounded everywhere, this implies that each term at the right--hand
side of (\ref{eq:curvIV}) is bounded as well. In particular, the
bounded character of the second term on the right--hand side of
(\ref{eq:curvIV}) guarantees that $\varphi^2\neq1$. On the one
hand, the domain of outer communications $\ll\!\!\mycal{I}\!\!\gg$
is a connected set, hence, its image under the continuous function
$\varphi$ is also a connected set, which implies we have only one
of the following two cases: or $\varphi^2<1$ in the whole of
$\ll\!\!\mycal{I}\!\!\gg$, or inversely $\varphi^2>1$ in all of
$\ll\!\!\mycal{I}\!\!\gg$. On the other hand, asymptotic flatness
($R=0$ at infinity) demands that each term at the right--hand side
of (\ref{eq:curvIV}) vanishes asymptotically, in particular,
vanishing of the second term implies that $\varphi$ approaches the
zero value at infinity, i.e., only the range
\begin{equation}
\varphi^2<1,
\label{eq:<1}
\end{equation}
is allowed for the scalar field in the domain of outer
communications $\ll\!\!\mycal{I}\!\!\gg$ of a static
asymptotically flat black hole.

Substituting the scalar curvature from (\ref{eq:curvIV}) into the
scalar equation (\ref{eq:K-G}), one obtains
\begin{equation}
\Box\varphi =\frac{\varphi}{1-\varphi^2}\left(
\nabla_\mu\varphi\nabla^\mu\varphi +\lambda{b}^2\varphi^2\right).
\label{eq:scalarIV}
\end{equation}

The rest of the proof of vanishing of the scalar field $\varphi$
in all of $\ll\!\!\mycal{I}\!\!\gg$ resembles the proof previously
accomplished in Sec.~\ref{sec:les}. Multiplying the scalar
equation (\ref{eq:scalarIV}) by $\varphi$, and integrating by
parts in the volume $\mycal{V}$ using the Gauss theorem yields
this time the integral identity
\begin{equation}
\left[\int_{\Sigma^{\prime}} -\int_\Sigma +\int_{\mycal{H}^+\cap
\overline{\mycal{V}}} +\int_{i^o\cap\overline{\mycal{V}}}\right]
\varphi\nabla^\mu\varphi\,d\Sigma_\mu
=\int_\mycal{V}\left(
\frac{\nabla_\mu\varphi\nabla^\mu\varphi}{1-\varphi^2}
+\lambda{b}^2\frac{\varphi^4}{1-\varphi^2}\right) dv.
\label{eq:intIV}
\end{equation}
The boundary integral in (\ref{eq:intIV}) is identical to the one
appearing in Sec.~\ref{sec:les}, and $\varphi$ satisfies similar
properties to those satisfied by the redefined scalar field of the
just mentioned section, i.e., $\varphi$ vanishes asymptotically
with suitable fall--off, and it is a bounded function at the
horizon (\ref{eq:<1}). All of the above implies that there are no
nonzero contributions in the boundary integral on the left--hand
side of (\ref{eq:intIV}), from which one concludes that
\begin{equation}
\int_\mycal{V}\left(
\frac{\gamma_{ij}\nabla^i\varphi\nabla^j\varphi}{1-\varphi^2}
+\lambda{b}^2\frac{\varphi^4}{1-\varphi^2}\right)dv=0,
\label{eq:zerosIV}
\end{equation}
in the coordinates of (\ref{eq:static}). Due to the condition
(\ref{eq:<1}) and since $\bm{\gamma}$ is positive definite, each
term in the integrand above is non--negative. Hence, vanishing of
the volume integral implies vanishing of each one of these terms,
and in particular, vanishing of the scalar field $\varphi$ in the
volume $\mycal{V}$, result which can be extended to the rest of
the domain of outer communications $\ll\!\!\mycal{I}\!\!\gg$ by
the staticity of the scalar field.

\section{Conclusions}
\label{sec:conclu}

We have shown the following ``no--scalar--hair'' theorem: for any
value of the nonminimal coupling parameter $\zeta$, a
self--gravitating scalar field with quartic self--interaction and
nonminimally coupled to gravity unavoidably vanishes in the domain
of outer communications of a static asymptotically flat black
hole. In particular, for the conformal coupling $\zeta=1/6$ the
conformal invariance of the system has no relevance in the
existence of nontrivial black hole solutions, as it would be the
case in the absence of self--interaction. The established results
completely settle the Bekenstein ``no--scalar--hair''
conjecture~\cite{Bekens96}, on the behavior of scalar systems in
presence of black holes, for quartic self--interactions and any
value of the nonminimal coupling parameter. With a vanishing
scalar field the right--hand side of the Einstein equations
(\ref{eq:Ein}) is zero, and the black hole associated to the
system is necessarily the Schwarzschild one.

\begin{acknowledgments}
The author thanks Alberto Garc\'{\i}a for helpful discussions and
support, and the staff of the Physics Department at CINVESTAV also
for support. This research was partially supported by the CONACyT
Grants 38495E and 34222E, together with the CONICYT Grant
FONDECYT-1010485, and the CONICYT/CONACyT Grant 2001-5-02-159. The
author thanks Isabel Negrete for typing the manuscript.
\end{acknowledgments}


\begin{thebibliography}{}

\bibitem{Bizon94} P. Bizo\'n,
Acta Phys. Polon. \textbf{B25}, 877 (1994).

\bibitem{Galtsov01} D.V. Gal'tsov,
``Gravitating Lumps,'' \texttt{hep-th/0112038} (2001).

\bibitem{Bekens96} J.D. Bekenstein,
%``Black Hole Hair: Twenty--Five Years After,''
%\texttt{gr-qc/9605059},
in \emph{Proceedings of Second Sakharov Conference in Physics,
Moscow}, edited by I.M. Dremin, A.M. Semikhatov (Singapore, World
Scientific, 1997).

\bibitem{Sawyer77} R.F. Sawyer,
%``Possibility of a Static Scalar Field in the Schwarzchild Geometry'',
Phys. Rev. D \textbf{15}, 1427 (1977).

\bibitem{Brum78} B.E. Brumbaugh,
%``Nonlinear Scalar Field Dynamics in the Schwarzchild Geometry'',
Phys. Rev. D \textbf{18}, 1335 (1978).

\bibitem{HeuslerStrau92} M. Heusler and N. Straumann, Class. Quantum
Grav. \textbf{9}, 2177 (1992).

\bibitem{Heusler92} M. Heusler,
%``A ``No--Hair'' Theorem for Self--Gravitating Nonlinear Sigma Models'',
J. Math. Phys. \textbf{33}, 3497 (1992).

\bibitem{Sudarsky95} D. Sudarsky,
%``A Simple Proof of a ``No--Hair'' Theorem in Einstein--Higgs Theory'',
Class. Quantum Grav. \textbf{12}, 579 (1995).

\bibitem{Heusler95} M. Heusler,
Class. Quantum Grav. \textbf{12}, 779 (1995).

\bibitem{Bekens95} J.D. Bekenstein, Phys. Rev. D \textbf{51}, R6608 (1995).

\bibitem{MayoBeke96} A.E. Mayo and J.D. Bekenstein,
Phys. Rev. D \textbf{54}, 5059 (1996).

\bibitem{SaaPRD96}  A. Saa, Phys. Rev. D \textbf{53}, 7377 (1996).

\bibitem{AyonGMP00}  E. Ay\'on--Beato, A. Garc\'{\i}a, A. Mac\'{\i}as,
and J.M. P\'erez--S\'anchez,
%``Note on Scalar Fields nonminimally Coupled to $(2+1)$--Gravity,''
Phys. Lett. \textbf{495B}, 164 (2000).

\bibitem{PenaSudarsky01} I. Pe\~na and D. Sudarsky,
%``Numerical Evidence Against Black Holes with
%nonminimally Coupled Scalar Hair,''
Class. Quantum Grav. \textbf{18}, 1461 (2001).

\bibitem{XanZan91} B.C. Xanthopoulos and T. Zannias,
J. Math. Phys. \textbf{32}, 1875 (1991).

\bibitem{XanDia92} B.C. Xanthopoulos and T.E. Dialynas,
J. Math. Phys. \textbf{33}, 1463 (1992).

\bibitem{Klimcik93} C. Klimcik, J. Math. Phys. \textbf{34}, 1914 (1993).

\bibitem{Zannias95} T. Zannias,
%``Black Holes Cannot Support Conformal Scalar Hair,''
%preprint \texttt{gr-qc/9409030},
J. Math. Phys. \textbf{36}, 6970 (1995).

\bibitem{BBM70} N. Bocharova, K. Bronnikov, and V. Melnikov,
Vestn. Mosk. Univ. Fiz. Astron. \textbf{6}, 706 (1970).

\bibitem{Bekens74} J.D. Bekenstein, Ann. Phys. (NY) \textbf{82}, 535 (1974).

\bibitem{SudaZann98} D. Sudarsky and T. Zannias,
%``Spherical black holes cannot support scalar hair,''
%\texttt{gr-qc/9712083}
Phys. Rev. D \textbf{58}, 087502 (1998).

\bibitem{Bekens98} J.D. Bekenstein, ``Black Holes: Classical Properties,
Thermodynamics and Heuristic Quantization,'' \emph{Proceedings of
the 9th Brazilian School of Cosmology and Gravitation}, Rio de
Janeiro, Brazil, \texttt{gr-qc/9808028} (1998).

\bibitem{Ayon01} E. Ay\'{o}n--Beato,
in \emph{Exact Solutions and Scalar Fields in Gravity:  Recent
Developments}, edited by A. Mac\'{\i}as, J.L. Cervantes--Cota, C.
L\"ammerzahl (Kluwer Academic/Plenum Publishers, New York 2001).

\bibitem{ChrWald95} P.T. Chru\'sciel and R.M. Wald,
%``On the Topology of Stationary Black Holes'',
Class. Quantum Grav. \textbf{11}, L147 (1994).

\bibitem{Galloway95} G.J. Galloway,
%``On the Topology of the Domain of Outer Communication'',
Class. Quantum Grav. \textbf{12}, L99 (1995).

\bibitem{Carter87} B. Carter,
%``Mathematical Foundations of the Theory of
%Relativistic Stellar and Black Holes Configurations'',
in \emph{Gravitation in Astrophysics (Carg\`ese Summer School
1986)}, edited by B. Carter, J.B. Hartle (Plenum, New York 1987).

\bibitem{Zannias98} T. Zannias, J. Math. Phys. \textbf{39}, 6651 (1998).

\bibitem{Ayon02a} E. Ay\'on--Beato,
%``Improving the ``No--Hair'' Theorem for the Proca Field,''
in \emph{Recent Developments in Mathematical and Experimental
Physics Vol. A: Cosmology and Gravitation}, edited by A.
Mac\'{\i}as, F. Uribe, E. Diaz (Kluwer Academic/Plenum Publishers,
New York 2002).

\bibitem{AyonPerez01} E. Ay\'on--Beato and J.M. P\'erez--S\'anchez,
``No--Scalar--Hair Theorem for Conformally Coupled Massive
Fields,'' submitted to Phys. Lett. \textbf{B} (2002).

\end{thebibliography}
\end{document}